\PassOptionsToPackage{doi=false,isbn=false,url=false}{bibtex}
\documentclass[letterpaper, 10 pt, conference]{ieeeconf} 

\IEEEoverridecommandlockouts                              

\usepackage{graphicx} 
\graphicspath{ {./figures/} }
\usepackage{amsmath} 
\usepackage{amssymb}
\usepackage{algorithm}
\usepackage[noend]{algpseudocode}
\usepackage{xcolor}
\definecolor{tumblue}{RGB}{0,101,189}

\usepackage{stfloats}
\usepackage{booktabs}
\usepackage{setspace}
\usepackage{tabularx}
\usepackage{booktabs}
\usepackage[T1]{fontenc}
\usepackage[scaled=0.85]{beramono}
\usepackage{listings}
\usepackage{caption}
\usepackage{cleveref}
\usepackage{url}
\let\labelindent\relax
\usepackage{enumitem}

\Crefname{lstlisting}{Listing}{Listings}
\lstdefinestyle{sparqlstyle}{
  language=OCL,
  basicstyle=\footnotesize,
  stepnumber=1,
  numbersep=10pt,
  tabsize=2,
  showspaces=false,
  breaklines=true
}

\usepackage[nonumberlist,nopostdot,acronym,toc]{glossaries}
\newacronym{cca}{CCA}{Central Controller Agent}
\newacronym{llm}{LLM}{Large Language Model}
\newacronym{mas}{MAS}{Multi-Agent System}
\newacronym{pa}{PA}{Product Agent}
\newacronym{ra}{RA}{Resource Agent}
\newacronym{api}{API}{Application Programming Interface}
\newacronym{sme}{SME}{Subject Matter Expert}
\newacronym{fsm}{FSM}{Finite State Machine}
\newacronym{RepastS}{RepastS}{Repast Symphony}
\newacronym{CoT}{CoT}{Chain-of-Thought}
\newacronym{mtbf}{MBTF}{Mean Time Between Failures}
\newacronym{mttr}{MTTR}{Mean Time To Repair}
\newacronym{rag}{RAG}{Retrieval Augmented Generation}
\newacronym{agv}{AGV}{Automated Guided Vehicle}

\usepackage{balance} 
\usepackage{placeins}

\makeatletter
\newcommand\fs@betterruled{%
  \def\@fs@cfont{\bfseries}\let\@fs@capt\floatc@ruled
  \def\@fs@pre{\vspace*{5pt}\hrule height.8pt depth0pt \kern2pt}%
  \def\@fs@post{\kern2pt\hrule\relax}%
  \def\@fs@mid{\kern2pt\hrule\kern2pt}%
  \let\@fs@iftopcapt\iftrue}
\floatstyle{betterruled}
\restylefloat{algorithm}
\makeatother

\author{Jonghan Lim$^{1}$ and Ilya Kovalenko$^{2}$
\thanks{$^{1}$Jonghan Lim is with the Department of Industrial and Manufacturing, Pennsylvania State University, State College, USA
        (e-mail: jxl567@psu.edu).
        }%
\thanks{$^{2}$ Ilya Kovalenko is with the Department of Industrial and Manufacturing and the Department of Mechanical Engineering, Pennsylvania State University, State College, USA
        (e-mail: iqk5135@psu.edu).
        }%
}

\title{\LARGE \bf
A Large Language Model-Enabled Control Architecture for Dynamic Resource Capability Exploration in Multi-Agent Manufacturing Systems
}

\begin{document}
\setlength{\textfloatsep}{2pt}

\maketitle
\thispagestyle{empty}
\pagestyle{empty}

\begin{abstract}

Manufacturing environments are becoming more complex and unpredictable due to factors such as demand variations and shorter product lifespans. This complexity requires real-time decision-making and adaptation to disruptions. Traditional control approaches highlight the need for advanced control strategies capable of overcoming unforeseen challenges, as they demonstrate limitations in responsiveness within dynamic industrial settings. Multi-agent systems address these challenges through decentralization of decision-making, enabling systems to respond dynamically to operational changes. However, current multi-agent systems encounter challenges related to real-time adaptation, context-aware decision-making, and the dynamic exploration of resource capabilities. Large language models provide the possibility to overcome these limitations through context-aware decision-making capabilities. This paper introduces a large language model-enabled control architecture for multi-agent manufacturing systems to dynamically explore resource capabilities in response to real-time disruptions. A simulation-based case study demonstrates that the proposed architecture improves system resilience and flexibility. The case study findings show improved throughput and efficient resource utilization compared to existing approaches.

\end{abstract}

\section{Introduction}
\label{sec:introduction}

Modern manufacturing systems face increasing complexity due to unforeseen operational disruptions (e.g. equipment failures, unexpected demand shifts)~\cite{koren2018reconfigurable}. To keep competitiveness, manufacturers have been integrating advanced automation and control systems to increase adaptability~\cite{xu2018industry}. However, many existing control approaches are rigid, unable to adapt to real-time variability on the shop floor, which causes bottlenecks and delays. To improve adaptability,~\gls{mas} has been proposed to enable the coordination of products and resources~\cite{leitao2009agent}. By distributing decision-making across agents representing different components, such as products, machines, and robots, \gls{mas} strategies enhance flexibility in manufacturing~\cite{leitao2012past}.

While \glspl{mas} enhance manufacturing flexibility, significant challenges exist. One challenge is the lack of real-time decision-making for dynamic reconfiguration. Manufacturing settings produce vast amounts of dynamic data from systems, products, and resources. Most systems cannot process this data in real-time for decision-making~\cite{pulikottil2023agent}. Therefore, in \gls{mas}, an adaptive control architecture is needed to respond to unanticipated disturbances~\textit{(RCh1)}. Moreover, limited contextual awareness of~\gls{mas} presents another challenge. Contextual information, such as resource locations and process parameters, is processed by predefined rules and algorithms in manufacturing~\cite{pulikottil2023agent}. These rule-based models limit the ability to flexibly adapt control strategies and interpret complex contextual changes (RCh2). Hence,~\gls{mas} face challenges in improving operational efficiency based on real-time shop floor conditions. Agents also operate within certain defined ranges of capabilities determined at the initial state. This constraint limits agents utilizing wider capabilities, preventing them from adapting to changing situations~\cite{kovalenko2022toward}. Most systems require \gls{sme} to manually change these boundaries~\textit{(RCh3)}. Dealing with these challenges requires a framework that allows distributed agents to learn and explore wider capabilities~\cite{kovalenko2022toward}.

\begin{figure}[t]
\smallskip
\smallskip
    \captionsetup{belowskip=-1pt}
    \includegraphics[width=.48\textwidth]{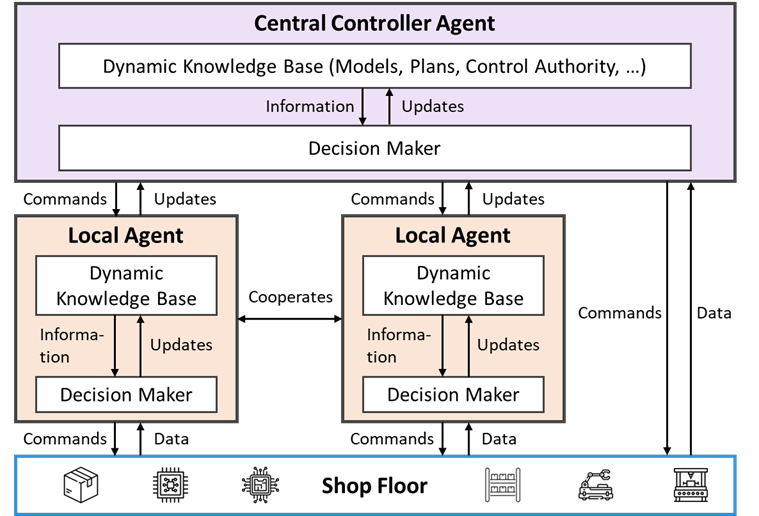}
    \caption{Overview of a control architecture that consists of multiple local agents including a central controller agent for manufacturing systems~\cite{kovalenko2022toward}}
    \label{fig:overview}
\end{figure}

\glspl{llm}, such as GPT-4o~\cite{hurst2024gpt} and Llama~\cite{touvron2023llama}, present opportunities to enhance adaptability. \glspl{llm} can support different tasks on the shop floor, enabling more effective communication. Previous studies have shown how \glspl{llm} can increase system flexibility and facilitate manufacturing interactions~\cite{lim2024large, lim2024enhancing}. \glspl{llm} have demonstrated the ability to interpret real-time data, support human-robot collaboration, and adapt to evolving operational requirements. Using these strengths, \glspl{llm} can be integrated into~\gls{cca} framework for~\gls{mas}, previously introduced in~\cite{kovalenko2022toward}, as presented in~\Cref{fig:overview}. This integration can improve real-time adaptability and dynamic exploration of resource capabilities. 

This paper presents an~\gls{llm}-enabled control architecture for~\glspl{mas}, enabling dynamic exploration of resource capabilities to adapt to real-time disruptions. The major contribution of this paper include:
(1) a control architecture that enables dynamic decision-making for unexpected disruptions,
(2) a context-aware system that evaluates the current state of the environment and performance to make informed control decisions, and
(3) a novel mechanism for dynamically exploring resource capabilities, improving agent coordination for system resilience.

The remainder of this paper is organized as follows. 
Section~\ref{sec:background} reviews control approaches and recent applications of~\glspl{llm} in manufacturing for adaptation.
Section~\ref{sec:overview} provides an overview of \gls{mas} and defines the problem.
Section~\ref{sec:architecture} describes the design of the control framework. 
Section~\ref{sec:casestudy} demonstrates the proposed framework through a case study. 
Finally, Section~\ref{sec:conclusion} summarizes the paper and suggests directions for future research.

\section{Background}
\label{sec:background}

\subsection{Dynamic Adaptation for Manufacturing System Control}
\label{subsec:control-strategies}

Different control strategies have been designed to enhance adaptability in manufacturing. A centralized control approach is used for resource allocation and scheduling~\cite{lopez2018software, qamsane2019dynamic}, leveraging sensor data for decision-making and rerouting parts during disruptions. However, as system complexity grows, centralized architectures face challenges in learning new information as they depend on predefined capabilities~\cite{kovalenko2022toward}. Distributed control strategies, such as~\gls{mas}, have been introduced to address these challenges, where various agents make local decisions under coordination to fulfill global objectives~\cite{leitao2009agent, leitao2012past}. Recent~\gls{mas} architectures integrate~\glspl{pa} and~\glspl{ra} to enhance flexibility~\cite{kovalenko2019model}. \glspl{pa} manage product-specific decision-making, while~\glspl{ra} control shop-floor resources like robots and machines~\cite{kovalenko2019model}. Existing~\gls{mas} solutions improve adaptability by \glspl{pa} dynamically exploring~\glspl{ra} for task execution~\cite{kovalenko2019dynamic}, while~\glspl{ra} optimize resource allocation during disruptions~\cite{bi2024dynamic}.

Although these systems improve adaptability, they remain limited in their initial configurations. These approaches do not address extending capabilities outside of specified limits~\cite{kovalenko2022toward}. If a resource is initially configured with a subset of its possible functions, existing systems can only adapt within these predefined capabilities. Furthermore,~\gls{mas} lacks a global perspective, where local adaptations are done without a holistic view. This paper introduces a~\gls{cca} within~\gls{mas}, leveraging an~\gls{llm} to assess real-time system conditions, learn from disturbances, and dynamically extend capabilities beyond initial configurations.

\subsection{Large Language Models in Manufacturing for Adaptation}
\label{subsec:control-strategies}

The introduction of~\glspl{llm} has provided possibilities for enhancing adaptability across various aspects of manufacturing. Makatura et al.~\cite{makatura2023can} and Li et al.~\cite{li2024large} demonstrate how~\gls{llm} translate natural language instructions into executable tasks, supporting flexible process planning and adaptive CAD/CAM workflows. By error-assisted fine-tuning and intelligent retrieval, Xia et al.~\cite{xia2024leveraging} and Freire et al.~\cite{kernan2024knowledge} further improve domain-specific programming and knowledge sharing. Garcia et al.~\cite{Garcia_DiBattista_Letelier_Halloran_Camelio_2024} and Fan et al.~\cite{Fan_Liu_Fuh_Lu_Li_2025} use~\gls{llm} for data processing and decision-making, allowing real-time task adjustments based on production constraints. Ni et al.~\cite{Ni_Wang_Leng_Chen_Cheng_2025} investigate~\gls{llm}-based process planning, highlighting~\glspl{llm}' role in increasing adaptability across manufacturing processes.

While these studies highlight the adaptability of~\glspl{llm} in manufacturing, they focus on simplified processes such as translating basic instructions, process planning, and knowledge retrieval. They do not focus on large-scale manufacturing systems, particularly in multi-agent environments where interactions between agents affect system performance. A gap exists in utilizing~\glspl{llm} for context-aware decision-making and exploring resource capabilities in response to unexpected disruptions. For instance, if multiple machines operate with varying efficiency levels,~\gls{llm} should not only retrieve performance data but also dynamically assess current states and constraints to enable reconfigurations. Therefore, we introduce a control architecture that utilizes~\glspl{llm} to explore capabilities and enable dynamic decision-making in large-scale manufacturing operations.

\section{Overview and Problem Statement}
\label{sec:overview}

\begin{figure*}[t]
\smallskip
\smallskip
    \centering
    \captionsetup{belowskip=-15pt}
    \includegraphics[width=.93\textwidth]{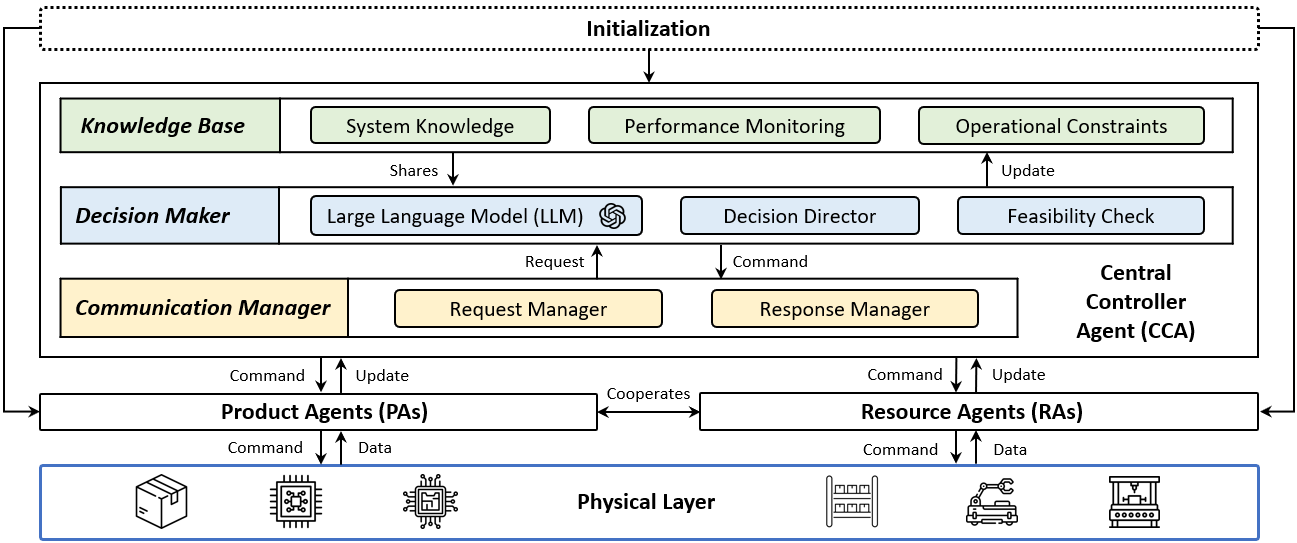}
    \caption{A Framework of Central Controller Agent for Multi-Agent Manufacturing Systems}
    \label{fig:architecture}
\end{figure*}

This section provides an overview of~\gls{mas}, which utilizes~\glspl{pa} and~\glspl{ra}, and outlines the resource capability exploration problem.

\subsection{Overview of Multi-Agent Manufacturing Systems}
\label{subsec:overview}

The majority of \glspl{mas} include \glspl{pa} and \glspl{ra}~\cite{kovalenko2019model}.~\glspl{pa} are responsible for physical parts, making decisions about processing sequences~\cite{kovalenko2019model}. \glspl{ra} manage and control physical resources, such as machines or robots, executing tasks requested by~\glspl{pa}~\cite{kovalenko2019dynamic}.~\glspl{pa} coordinate with~\glspl{ra} to explore available resource capabilities and make informed decisions for guiding physical parts~\cite{kovalenko2019dynamic}.

\subsubsection{Resource Agents}
\label{subsubsec:capabilities}

Each \gls{ra} is equipped with a~\gls{fsm} model that specifies its capabilities and interactions with neighboring~\glspl{ra}, which are \glspl{ra} that are sharing common states for coordinated operations. The capabilities model, $M$, is formally defined for~\gls{ra} and contains the following elements~\cite{kovalenko2019dynamic}:

\noindent 
$M = (X, E, Tr, Prp_p, x_i, X_m)$:

\begin{itemize}[leftmargin=*]
    \item[] $X = \{x_0, x_1, \dots, x_n\}$ a set of states that a physical part can achieve when processed by the resource        
    \item[] $E = \{e_0, e_1, \dots, e_l\}$ is a set of events that the \gls{ra} can start, each corresponding to a process in the physical system
    \item[] $Tr: X \times E \rightarrow X$ a function that maps state-event pairs to resulting states
    \item[] $Prp_p: X \rightarrow P_p$ a function that maps each state to physical properties that change the part
    \item[] $x_i \in X$: the initial state of the part when the resources evaluate a bid request
    \item[] $X_m \subseteq X$: a set of states that the part must reach to complete a task, updated with each bid request
\end{itemize}
\glspl{ra} update their capability models using shop floor data and agents~\cite{kovalenko2019dynamic}. \glspl{ra} use shared states \(X_s \subseteq X\) to communicate, representing conditions accessible to multiple resources. Each \gls{ra} maintains a table \(N: X_s \rightarrow 2^{RA}\), which allows~\glspl{pa} to forward task execution requests to neighboring~\glspl{ra} for task allocation~\cite{kovalenko2019dynamic}.

\subsubsection{Product Agents}
\label{subsubsec:product_agent}

The~\gls{pa} utilizes three main components to support decision-making and interaction with the~\gls{ra}: the process plan, product history, and environment model. Additional details and examples of \gls{ra} and \gls{pa} models are provided in~\cite{kovalenko2019model}.

The process plan represents a customer order as an ordered list of physical property sets: $P_d = (P_{d1}, P_{d2}, \dots, P_{dn})$. Each element \(p_{ij} \in P_p \cup P_{np}\) defines a desired property, where \(P_p\) changes the part’s composition, and \(P_{np}\) does not (e.g., location, orientation). The plan is sequential, requiring completion of \(P_1\) before \(P_2\), and so on.

The product history represents the current state and progression of a part. Its definition is:  $PH = (X_v, Prp_v, RA_c)$, where $X_v = \{x_0, x_1, \dots, x_m\}$ denotes the sequence of achieved states, $Prp_v: X_v \to P_p \cup P_{np}$ represents the set of completed physical properties, and $RA_{c}$ is the \gls{ra} that informed the \gls{pa} of a visited state. As the product moves through different processes, each \gls{ra} updates the \gls{pa} with information about state transitions and physical properties.

The environment model $M_e$ provides \gls{pa} with a holistic view of the manufacturing environment. The environment model $M_e$ is defined as: $M_e = (X_e, E_e, Tr_e, Prp_{p,e}, Ag_{h,e}, x_c)$, where $x_c$ is the current state of the physical part. $Ag_{h,e}$ is the event-agent association function, mapping each event to the responsible \gls{ra}. The state set $X_e$, event set $E_e$, state transition function $Tr_e$, and physical properties $Prp_{p,e}$ follow the same definitions in the capabilities model $M$.

\subsection{Problem Statement}
\label{subsec:problem}

The dynamic capability exploration problem is the process of adapting an~\gls{ra} in response to disruptions by extending its operational capabilities beyond predefined boundaries. When an~\gls{ra} cannot fulfill its assigned tasks, the system must identify an exploration agent $RA_e$, that can take over the disrupted operations. This involves reconfiguring capabilities, e.g. extending a robot’s reach or adjusting a CNC spindle speed if possible or needed. To formally define this problem, we introduce the following:
\begin{equation}
\mathcal{F}(RA_d, X_{d}, E_d) = (RA_e, C_e) \rightarrow (X_e, E_e)
\end{equation}
where $RA_d$ is the disrupted agent unable to perform its operations, and $X_{d}$ and $E_{d}$ represent the initial states and events before the disruption. The function $\mathcal{F}$ identifies an exploration agent $RA_e$ and its explored capabilities $C_e$ (e.g. movement locations, process parameters). These capabilities define the set of feasible states, $X_e$, and the corresponding executable events, $E_e$. This formulation guarantees exploration agent, $RA_e$, can meet the required process plan $P_d$.

For the exploration agent, \(RA_e\), to take over disrupted tasks, each state transition must follow a sequence that support process completion. These valid state transitions ensure resource progress through the states to fulfill the assigned task. This requirement is expressed as:
\begin{equation}
\forall x_i \in X_e, \forall e_l \in E_e, \quad Tr(x_i, e_l) = x_j \land x_j \in X_m
\end{equation}
where $Tr(x_i, e_l) = x_j$ represents the state transition function, mapping a state-event pair to a new state. The condition $x_j \in X_m$ ensures that the resulting state $x_j$ belongs to the set of marked states, which represent states for task completion. This guarantees that new capabilities of the exploration agent $RA_e$ can achieve process plan $P_d$. To meet these requirements, this paper proposes a control architecture that enables dynamic resource capability exploration.

\section{Architecture}
\label{sec:architecture}

This section presents a \gls{cca} framework for~\gls{mas}, as depicted in~\Cref{fig:architecture}, which enables~\glspl{ra} to explore dynamic resource capabilities. The ~\gls{cca} framework consists of three main components: a knowledge base, a decision-maker, and a communication manager.

\subsection{Central Controller Agent Knowledge Base}
\label{subsec:cca-knowledge}

The knowledge base of~\gls{cca} gathers knowledge from \glspl{pa} and \glspl{ra}. The~\gls{cca} dynamically explores resource capabilities based on real-time data by utilizing information from local agents.

\subsubsection{System Knowledge}
\label{subsubsec:performance}

The \gls{cca} collects information from \glspl{ra} regarding their status, capabilities, and performance. This includes each \gls{ra}'s capabilities, current tasks indicating which product each \gls{ra} is processing, and monitors resource status to identify operational disruptions.
The \gls{cca} also gathers information from product history $PH$, process plan $P_d$, environment model $M_e$ to identify exploration agent $RA_e$. The product history $PH$ provides a record of achieved states $X_v$, the set of completed physical properties $Prp_v$, and the last resource agent $RA_c$ that informed the \gls{pa}.

\subsubsection{Performance Monitoring}
\label{subsubsec:performance}

By monitoring resource capacity, the \gls{cca} collects performance metrics to enhance decision-making. The~\gls{cca} gathers data at discrete time steps from \glspl{ra}, constructing a performance vector \(PM_i(t)\) that includes performance indicators such as throughput, utilization, breakdown status, and availability. This is defined as \(PM_i(t) = [m_1(t), m_2(t), \dots, m_n(t)]\), where each element \(m_j(t)\) represents a specific metric measured at discrete time \(t\). Continuous data updates guide the \gls{cca}'s informed decision with both current states and performance.

\subsubsection{Operational Constraints}
\label{subsubsec:authority}

The \gls{cca} maintains operational boundaries and safety constraints that determine the permitted limits within each \gls{ra}. These limits represent the thresholds for resource operation (e.g. motion range, load capacity) to guarantee feasible and safe exploration. Feasible exploration enables an~\gls{ra} to function within its boundaries without exceeding constraints, such as a robot arm operating within its workspace. Safe exploration prevents hazardous states, such as spindle speed or feed rate of a CNC remaining within safe limits to prevent tool breakage. The operational boundaries determine the range of allowable actions, while the safety constraints prevent hazardous behavior. The operational boundaries for a feasible exploration of a resource \(i\) can be represented as ${B}_i = \{ b_{i1}, b_{i2}, \dots, b_{in} \}$, where each element \( b_{ij} \) corresponds to a specific operational parameter with defined minimum and maximum limits $b_{ij} \in [b_{ij}^{\min}, b_{ij}^{\max}]$. Each resource is also subject to safety constraints to prevent unsafe states. The safety constraints for a safe exploration of a resource \(i\) can be represented as: $ {S}_i = \{ s_{i1}, s_{i2}, \dots, s_{ik} \}$, with defined minimum and maximum limits $s_{ik} \in [s_{ik}^{\min}, s_{ik}^{\max}]$. The feasibility of a resource executing an explored capability is determined using the validation function discussed in~\Cref{subsec:outputs}.

\subsection{Decision Maker}
\label{subsec:decision-maker}

\begin{figure}[t]
\smallskip
\smallskip
    \captionsetup{belowskip=-1pt}
    \includegraphics[width=.48\textwidth]{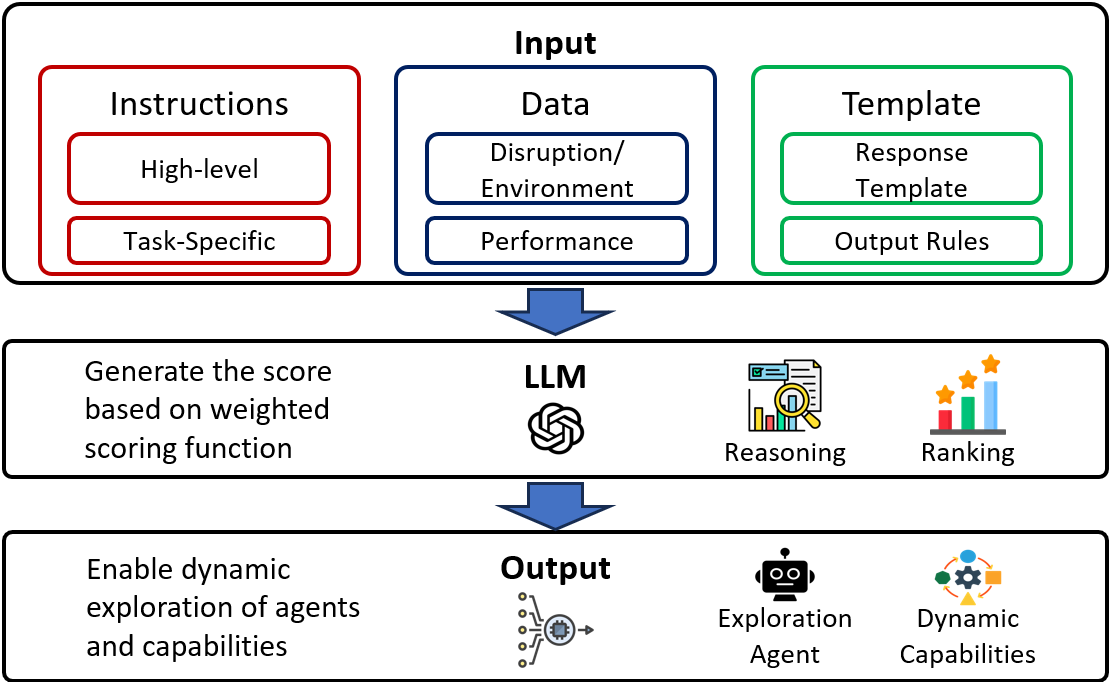}
    \caption{Interface for \gls{llm}-Enabled Decision-Making Process}
    \label{fig:llmInterface}
\end{figure}

Using the knowledge detailed in~\Cref{subsec:cca-knowledge}, the decision maker is responsible for authorizing agents with explored capabilities. \Cref{fig:llmInterface} illustrates the \gls{llm}-driven decision-making process, which consists of three key stages: input preparation, \gls{llm} for exploration, and output generation.

\subsubsection{Inputs}
\label{subsec:llm-interface}

\glspl{llm} enhance reasoning through rationale exploration, generating and evaluating multiple reasoning paths to improve problem-solving~\cite{huang2022towards}. To ensure structured and reliable outputs, we employ constraint-based prompts~\cite{lu2023bounding} and structured~\gls{CoT} prompts~\cite{wei2022chain} to enhance the transparency of~\gls{llm} decisions. Structured~\gls{CoT} prompts guide the~\gls{llm} to explicitly outline reasoning steps, clarifying why certain~\gls{ra} or capabilities are selected during disruptions. The decision-maker component utilizes three types of structured inputs for \gls{llm}:
\begin{itemize}[nosep]
 \item \textbf{Instructions: }Instructions guide the~\gls{llm} to analyze real-time data and explore resource capabilities while considering constraints, disruptions, and performance. 
 \item \textbf{Data: }The \gls{llm} receives disrupted agent \(RA_d\) and its capabilities, performance vectors \(PM_i(t)\), and product data including process plan $P_d$, product history $PH$, and environment model $M_e$.
 \item \textbf{Template: }The template structures the \gls{llm}'s output to ensure consistency so that the capabilities can be applied to exploration agent $RA_e$.
\end{itemize}

\subsubsection{\gls{llm} for Resource Capability Exploration}
\label{subsec:decision-director}

Various studies have applied ``\gls{llm}-as-a-judge'' across diverse applications for scoring, ranking, and selection~\cite{li2024generation}. Although rule-based methods (e.g. selecting the nearest robot) provide consistent decisions, they lack adaptability to unforeseen disruptions. Additionally, \gls{llm}-based decision-making strategies, such as unconstrained reasoning, lack systematic methods to ensure reproducibility. In this work, we leverage \glspl{llm} to evaluate \glspl{ra} for capability exploration using a weighted scoring function:
\vspace{-0.8em}

\begin{algorithm}
\caption{Exploration Resource Agent Validation}
\label{alg:validate_resource}
\begin{algorithmic}[1]
\algrenewcommand\algorithmicindent{0.7em}
\scriptsize{
\Procedure{validate}{$O = (RA_e, C_e)$, $exploration\_params$, $n$}
    \State $i \gets 0$, $valid \gets false$, $feedback \gets \text{None}$
    \While{$i < n$ \textbf{and} not $valid$}
        \State $i \gets i + 1$

        \If{\text{syntaxCheck(O) = false}}
            \State $feedback \gets \text{"Syntax error"}$
            \State \textbf{continue}
        \EndIf
        
        \State $RA_e \gets getInfo(O)$
        \If{$RA_e \notin model$}
            \State $feedback \gets \text{"Invalid agent"}$
            \State \textbf{continue}
        \EndIf
        
        \State $C_e \gets RA_e.capabilities$
        \State $(b_{ij}^{\min}, b_{ij}^{\max}) \gets exploration\_params.operation\_bounds$
        \State $(s_{ik}^{\min}, s_{ik}^{\max}) \gets exploration\_params.safety\_limit$
        
        \If{$\forall c_m \in C_e,$
            \quad $b_{ij}^{\min} \leq c_m \leq b_{ij}^{\max} \text{ and}$
            \quad $s_{ik}^{\min} \leq c_m \leq s_{ik}^{\max}$}
            \State $valid \gets true$
        \Else
            \State $feedback \gets \text{"Constraints not met"}$
        \EndIf
    \EndWhile
    \State \Return $valid, feedback$
\EndProcedure
}
\end{algorithmic}
\end{algorithm}

\begin{equation}
S_i = \sum_{j=1}^{n} \omega_j \cdot f_j(x_{ij})
\end{equation}
where score \(S_i\) represents the suitability of a candidate resource agent \(i\) for exploration based on multiple performance indicators. \(\omega_j\) is the weight assigned to factor $j$ by the \gls{llm}, and \(x_{ij}\) is the raw value, such as throughput, utilization, or breakdown rates. The function \(f_j\) normalizes these values. The agent with the highest score is selected as exploration agent \(RA_e\), and its capabilities \(C_e\) are adjusted by exploring the disrupted agent’s functional responsibilities. The scoring approach provides structured, quantifiable comparisons among candidates, aiding transparency and validation. In this study, the weights $\omega_j$ were determined by the~\gls{llm}. Systematic and prompt design for explicit weight determination will be done in future work to improve decision reliability.

\subsubsection{Outputs}
\label{subsec:outputs}

The output $O$ consists of the exploration agent \(RA_e\) and explored capabilities \(C_e\), which is represented as \(O = (RA_e, C_e)\). The \gls{cca} processes this output to reconfigure the resource’s capabilities model \(M\) with states $X$ and events $E$.

Although \glspl{llm} are effective for analyzing and generating information, their outputs can be inconsistent. Several studies, such as Self-Refine~\cite{madaan2023self}, have employed iterative refinement to reduce errors. We apply a similar approach, using feedback-driven adjustments to ensure \gls{llm} decisions remain within predefined operational and safety constraints. This method is critical in manufacturing contexts, where safety constraints and operational parameters are essential to maintain reliability.

The validation process, as demonstrated in \Cref{alg:validate_resource}, ensures that the selected agent and capabilities comply with both system constraints and safety limits.
The validation begins with the \gls{cca} retrieving the operational boundary set \(\mathcal{B}_i\) and safety constraint set \(\mathcal{S}_i\). Before proceeding, the generated output $O$ is checked for syntax error, ensuring the output $O$ is generated based on the expected structure. If an error is detected, feedback is generated and used to refine the next iteration. The \gls{cca} then initializes the validation loop, setting the iteration counter to zero and marking the output as invalid. 
For each iteration, the \gls{cca} extracts information of exploration agent \(RA_e\) from \(O = (RA_e, C_e)\). If \(RA_e\) is not found, the validation process continues with updated feedback. The corresponding explored capabilities, \(C_e\) of \(RA_e\) are then retrieved and compared against \(\mathcal{B}_i\) and \(\mathcal{S}_i\). If these conditions hold, the validation is successful, and the output is applied. Otherwise, feedback is provided, and the process continues until a valid response is obtained or the iteration limit \(n\) is reached.\vspace{-0.2em}

\begin{figure}[t]
\smallskip
\smallskip
    \captionsetup{belowskip=-5pt}
    \includegraphics[width=.48\textwidth]{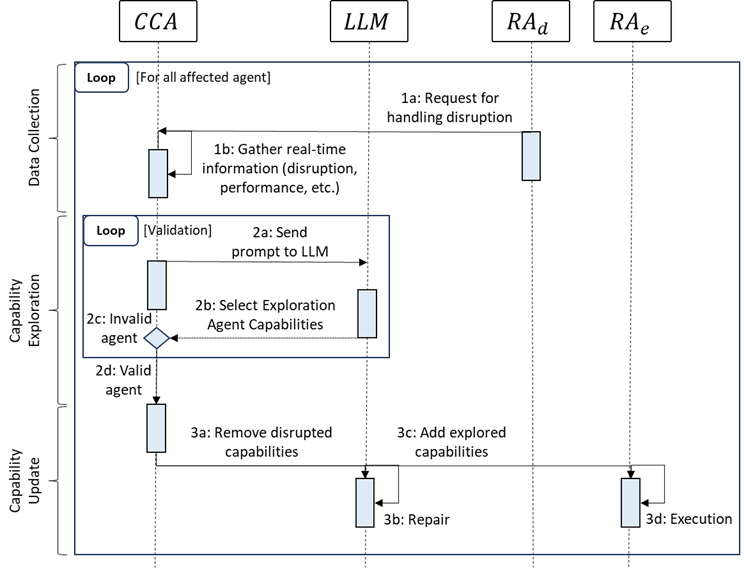}
    \caption{Sequence Diagram for Exploration Process}
    \label{fig:sequenceDiagram}
\end{figure}

\subsection{Communication Manager}
\label{subsec:communication}

The Communication Manager is responsible for updating the capabilities model $M$ based on the output \(O = (RA_e, C_e)\) received from the \gls{llm}. This process begins by extracting the original event set \(E^{\text{original}}\) and \(X_s\) from the selected \(RA_e\). Disrupted events from \(RA_d\) are removed to prevent invalid transitions.

The newly assigned \(C_e\) are converted into \(e_j \in E\), each defining specific state transitions. The event set \(E'\) is constructed by merging the original events with the newly assigned events, represented as: 
\vspace{-0.8em}

\begin{equation}
E' = E^{\text{original}} \cup E^{\text{new}}
\end{equation}
This integration enhances the~\gls{ra}’s operational efficiency by enabling more direct and flexible state transitions, as captured by the updated state transition function \(Tr\), which maps initial states \(x_i \in X\), to target states within the marked state \(X_m\).

Additionally, neighboring agents are added to the neighbor table \(N: X_s \rightarrow 2^{RA}\) for enhanced coordination. The~\glspl{pa} then cooperates with~\glspl{ra} through a bidding process to select the most efficient path while considering constraints and processing time~\cite{kovalenko2019dynamic}.

\subsection{Exploration Process Workflow}
\label{subsec:exploration}

The sequence diagram, as depicted in \Cref{fig:sequenceDiagram}, illustrates the exploration process. The process is divided into three primary stages: \textit{Data Collection}, \textit{Capability Exploration}, and \textit{Capability Update}.

The \textit{Data Collection} stage begins with \gls{cca} detecting a disruption and sends a request for handling the event to $RA_d$ (Step 1a). The \gls{cca} gathers real-time information on the disruption and the performance (Step 1b). This step is performed for all affected agents within a loop.

In the \textit{Capability Exploration} stage, the \gls{cca} sends a prompt to the \gls{llm} to handle disruption (Step 2a). The \gls{llm} then selects an exploration agent $RA_e$ and explored capabilities $C_e$ (Step 2b). The exploration agent $RA_e$ and explored capabilities $C_e$ undergo validation, where each capability is checked against predefined operational and safety constraints. If a capability does not meet the criteria, the agent is invalid (Step 2c), and the exploration loop continues. If all constraints are satisfied, the explored capabilities $RA_e$ handle disruption (Step 2d).

During the \textit{Capability Update}, the \gls{cca} removes disrupted capabilities from the capability list (Step 3a) and triggers the repair process (Step 3b). The explored capabilities $C_e$ are added to the exploration agent $RA_e$ (Step 3c), expanding its functionality. The exploration agent $RA_e$ then proceeds to execute its tasks using the updated capabilities (Step 3d) by collaborating with~\glspl{pa}.

\begin{figure*}[t]
\smallskip
\smallskip
    \centering
    \captionsetup{belowskip=-15pt}
    \includegraphics[width=.96\textwidth]{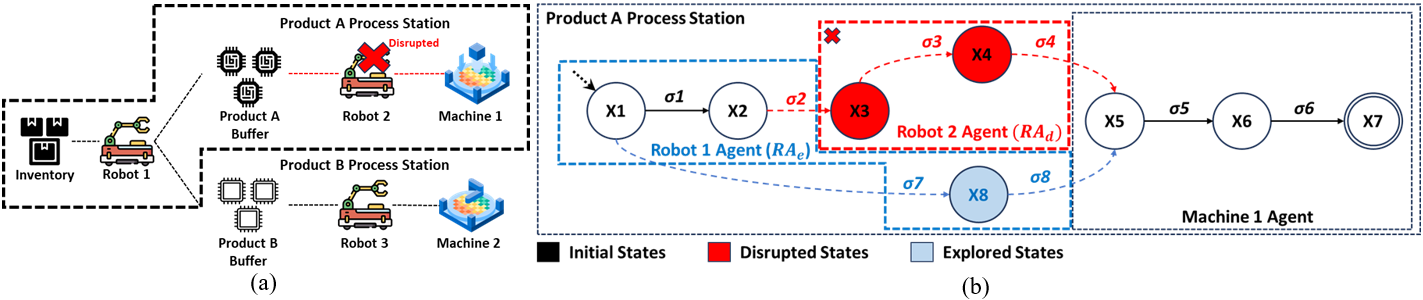}
    \caption{(a) An overview of a system with three robots, two machines, and three buffers. A disruption in Robot 2 affects Product A operations. (b) The \gls{cca} framework responds to the disruption at $X3$ and $X4$ by authorizing Robot 1 to expand its capabilities to take over the disrupted operations, ensuring task execution at $X8$.}
    \label{fig:example}
\end{figure*}

\subsection{Exploration Example}
\label{subsec:example}

\begin{figure}[t]
    \centering
    \captionsetup{belowskip=-3pt}
    \begin{minipage}{0.48\textwidth} 
        \centering
        \includegraphics[width=\textwidth]{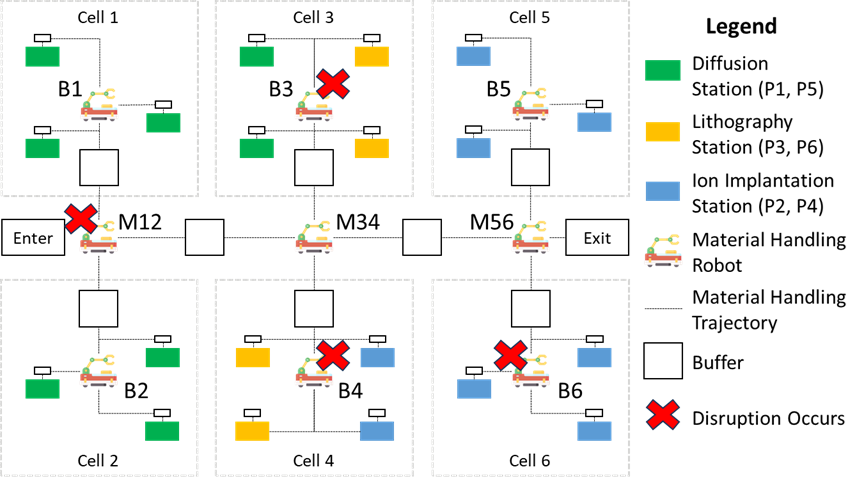}
        \caption{Case Study Setup}
        \label{fig:setup}
    \end{minipage}

    \vspace{0.3cm} 

    \begin{minipage}{0.48\textwidth} 
        \centering
        \captionof{table}{Machine Process Times}
        \label{tab:process-tick}
        \small 
        \begin{tabular}{|c|c|c|c|c|c|c|}
            \hline
            \textbf{Process} & P1  & P2  & P3  & P4  & P5  & P6  \\
            \hline
            \textbf{Time (Ticks)} & 150 & 60 & 110 & 100 & 170 & 20 \\
            \hline
        \end{tabular}
    \end{minipage}

    \vspace{0.3cm} 

    \begin{minipage}{0.48\textwidth} 
        \centering
        \captionof{table}{Breakdown Start Times and Recovery Times}
        \label{tab:robot-failures}
        \small 
        \begin{tabular}{|c|c|c|c|}
            \hline
            \textbf{Robot} & \textbf{Initial Breakdown (Tick)} & \textbf{MTTR (Ticks)} \\
            \hline
            M12 & 1000 & 450 \\
            B3 & 2500 & 450 \\
            B4 & 3000 & 340 \\
            B6 & 4500 & 390 \\
            \hline
        \end{tabular}
    \end{minipage}
\end{figure}

\Cref{fig:example} illustrates an example with three robots (Robot 1, Robot 2, Robot 3), two machines (Machine 1, Machine 2), and three buffers (Inventory, Product A Buffer, Product B Buffer). A disruption occurs at Robot 2, preventing it from transferring Product A to Machine 1, affecting states $X3$ (at Product A Buffer) and $X4$ (moving to Machine 1). The \gls{cca} collects \glspl{ra} status and performance \(PM_i(t)\) by retrieving data from~\gls{ra} through direct queries. Metrics, including availability (\(A_i\)), utilization (\(U_i\)), and proximity (\(P_i\)), are stored in a knowledge base for \gls{llm} to evaluate potential exploration agents \(RA_e\). For example, the user provides the following prompt:
\begin{quote}
{\noindent\scriptsize
\texttt{
"Analyze real-time data (R1: proximity=2, utilization=.92; R2: proximity=6, utilization=.35...) to select an exploration agent(availability, utilization, proximity). Enforce movement constraints and output your decision with clear reasoning..."}}
\end{quote}
Based on this instruction, the \gls{llm} calculates a score \(S_i\) using a weighted function $S_i = w_1 A_i + w_2 (1 - P_i) + w_3 U_i$ to select Robot 1 as the exploration agent \(RA_e\) over Robot 3, based on its higher score. The \gls{cca} then validates operational boundaries by checking movement range constraints, confirming that Robot 1 can execute the required operations. The \gls{cca} assigns Robot 1 as the $RA_e$ and updates its capability model by integrating the disrupted events $\sigma_3$ and $\sigma_4$. The event set $E'$ is modified, and shared states $X_s$ are updated, enabling Robot 1 to transition through $X8$ (Inventory to Machine 1) and reach $X5$ (at Machine 1). Through the bidding process, the~\gls{pa} identifies Robot 1 and state $X8$ as the most efficient path to achieve the marked state $X7$ (processed at Machine 1).

\section{Case Study}
\label{sec:casestudy}

To evaluate the \gls{cca} architecture, an~\gls{mas} was setup in a simulated environment. This section outlines the setup and showcases the exploration process in a scaled-up case study, followed by an analysis of the results.

\subsection{Case Study Setup}
\label{subsec:casestudy-setup}

The~\gls{RepastS}~\cite{north2013complex} platform is used to model and simulate~\gls{mas} within the \gls{cca} architecture. The platform updates at each time step, referred to as a tick~\cite{north2013complex}. This study modifies and extends previous works~\cite{kovalenko2019model, kovalenko2022cooperative}. The simulated facility, as shown in~\Cref{fig:setup}, consists of 20 stations (diffusion, ion implementation, and lithography) connected by a network of 9 material handling robots, each designated to its own cell or transfer stations. For example, robot B1 can only transfer parts within Cell 1, while robot M12 is responsible for transferring parts between Cell 1 and Cell 2. Buffers are positioned between the robots and also placed at each cell. Wafer lots enter the facility through an entrance and exit after completing a defined set of production steps. The facility supports six manufacturing processes (P1–P6), each performed at designated stations with processing times, as detailed in~\Cref{tab:process-tick}.
In this simulation, four robots experience disruptions starting with robot M12 at tick 1000. The initial breakdown and recovery times, including their~\gls{mttr}, are detailed in~\Cref{tab:robot-failures}. The \gls{cca} handles disruptions as outlined in~\Cref{sec:architecture}. In this case study, we employed GPT-4o by OpenAI~\cite{hurst2024gpt}.

\subsection{Case Study and Results}
\label{subsec:casestudy-results}

\begin{figure}[t]
\smallskip
\smallskip
    \captionsetup{belowskip=-15pt}
    \includegraphics[width=.48\textwidth]{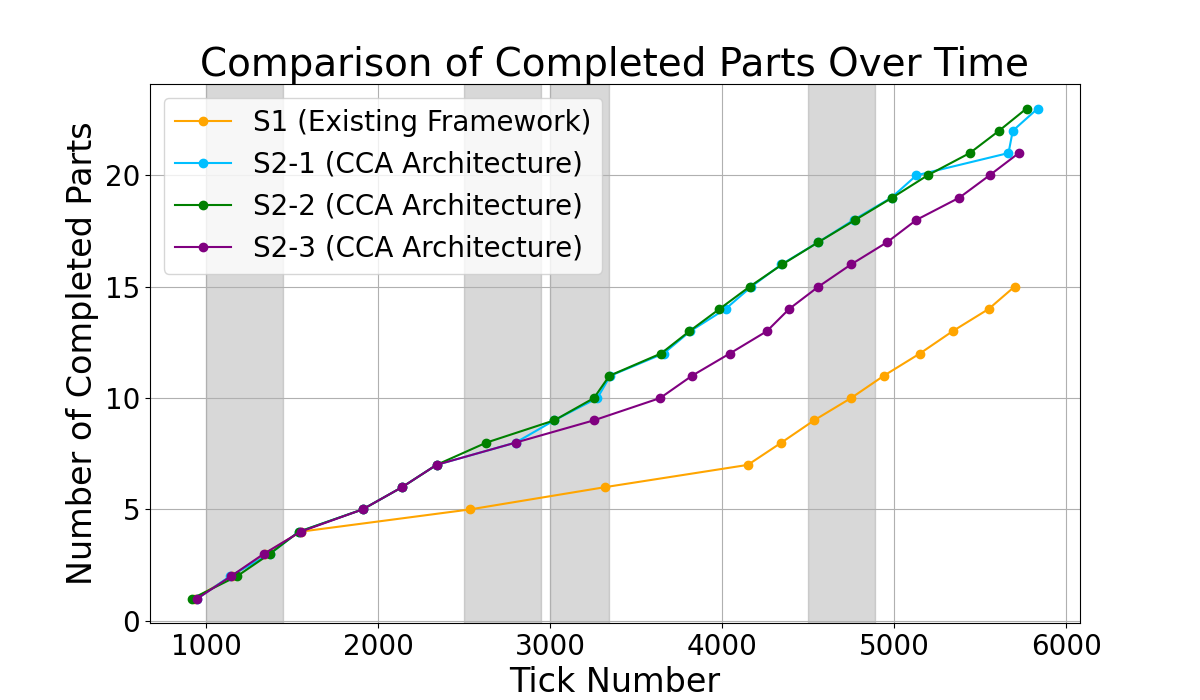}
    \caption{Throughput comparison between scenario S1 and three trials of scenario S2}
    \label{fig:throughput}
\end{figure}

\begin{figure}[t]
\smallskip
\smallskip
    \captionsetup{belowskip=-2pt}
    \includegraphics[width=.48\textwidth]{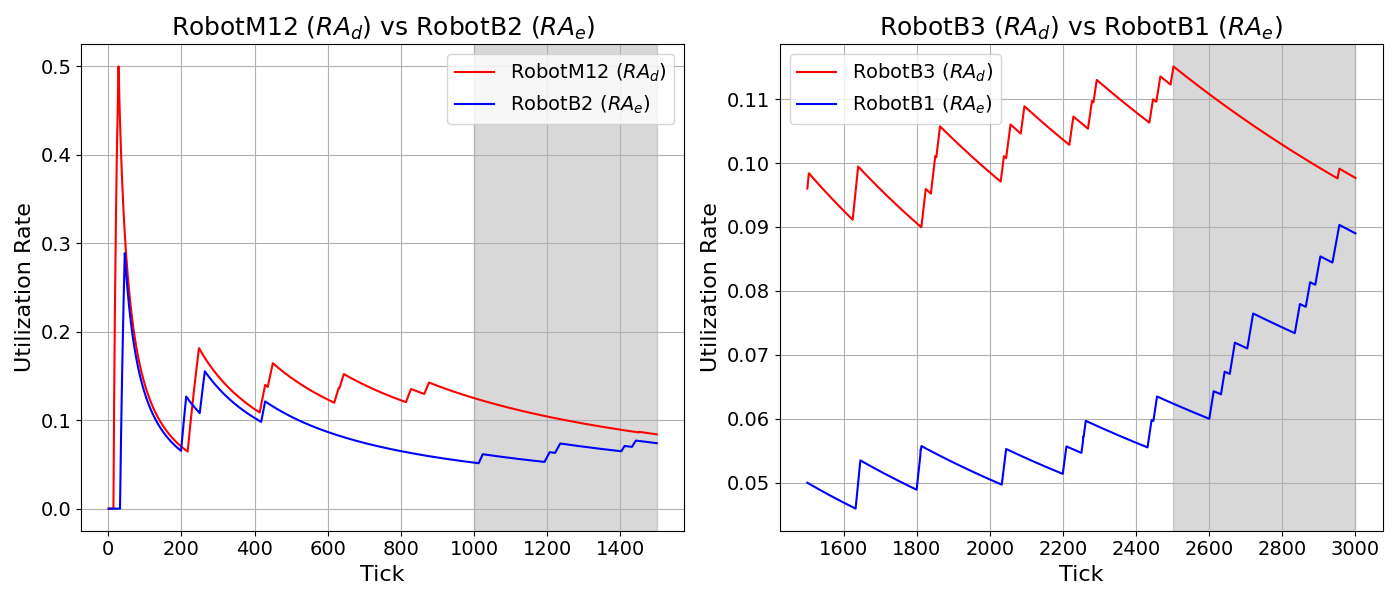}
    \caption{Utilization rate comparison between the exploration agent and the disrupted agent during breakdown in S2-3}
    \label{fig:utilization}
\end{figure}

A batch of 25 parts enters the system at tick 10, following the process sequence P1 $\to$ P2 $\to$ P3 $\to$ P4 $\to$ P5 $\to$ P6.
This study compares two scenarios: (S1) the existing framework of Kovalenko et al.~\cite{kovalenko2022cooperative}, and (S2) the \gls{cca} architecture, which was tested across three trials to mitigate disruptions. The performance was evaluated by comparing the number of completed parts over time and the utilization rates of $RA_d$ and $RA_e$.
The following assumptions are made for the case study:
\begin{list}{}{\setlength{\leftmargin}{0.7cm} \setlength{\labelwidth}{1.0cm} \setlength{\itemindent}{-0.0cm}}
    \item[A.1] The communication delay from \gls{llm} is not considered.
    \item[A.2] No two robots breaking down simultaneously.
    \item[A.3] Once restored, robots revert to their original capabilities.
\end{list}

A.1 ensures that decision-making performance is evaluated independently without interference from variable response times. A.2 prevents overlapping disruptions, allowing a clear evaluation of the system's response to individual failures. Finally, A.3 isolates the effects of dynamic capability exploration without additional variability.

The throughput results, shown in~\Cref{fig:throughput}, demonstrate a significant improvement with the \gls{cca} architecture in handling breakdowns, as indicated by the gray-shaded regions representing breakdown periods detailed in~\Cref{tab:robot-failures}. The existing framework (S1) completed only 15 parts, as disruptions prevented parts from finding feasible paths, leading to their removal. In contrast, the \gls{cca} architecture, tested in three trials S2-1, S2-2, and S2-3, mitigated failures, completing 23, 23, and 21 parts, respectively. Some parts still failed due to real-time disruptions, particularly when pre-planned routes became invalid or when parts were stuck in robots that had broken down, forcing their removal.

To further analyze the system behavior, the utilization rates of disrupted agent $RA_d$ and exploration agent $RA_e$ were compared, as observed in~\Cref{fig:utilization} for S2-3, where robot M12 and robot B3 experienced breakdowns. The \gls{cca} explored new capabilities for robot B2 and robot B1, respectively. The gray-shaded regions in~\Cref{fig:utilization} indicate the periods during which the robots experienced disruptions. During these periods, both the disrupted agents $RA_d$ and the exploration agents $RA_e$ exhibit similar behavior. $RA_d$ showed a decline in utilization and $RA_e$ demonstrated an increase as they were assigned new tasks.

\subsection{Insights from Case Study}
\label{subsec:casestudy-insights}

The case study highlights that the proposed \gls{cca} architecture enhances system resilience and adaptability. This research demonstrated how the system can maintain efficient operations despite disruptions, minimize part removals, and reduce failures. Even when agents experience breakdowns, the system sustains stable throughput and minimizes bottlenecks. These findings suggest that the~\gls{cca} architecture offers a robust and flexible approach for handling dynamic challenges.

However, there are several challenges identified in this study. A major issue is the limited explainability of~\gls{llm}-driven decision-making. Although agent exploration was based on a weighted scoring function, the~\gls{llm} did not always make the best decisions. Future research should focus on increasing prompt effectiveness and enhancing the explainability of~\glspl{llm}. For instance, the \gls{llm} occasionally assigned inconsistent weights to performance metrics, such as prioritizing utilization rate over proximity without clear justification. Therefore, reliable decision-making depends on refining the prompt design and increasing the transparency of \gls{llm} reasoning.

Another challenge is the generalizability of the proposed architecture. Although the system successfully handled unexpected robot breakdowns, further evaluation is needed to assess its applicability to other uncertainties, such as machine failures, demand shifts, and simultaneous disruptions of multiple agents. Although fine-tuning~\gls{llm} can enhance accuracy in manufacturing~\cite{xia2024leveraging} for specific scenarios, developing an approach capable of handling any disturbances is important for maintaining efficiency. Additionally, the accuracy of~\gls{llm} responses relies on structured and relevant runtime data. Therefore, the architecture should utilize a knowledge base that is designed for~\gls{mas} runtime control~\cite{lim2023ontology}. This can ensure the system retrieves relevant context related to real-time disturbances with~\gls{llm} using~\gls{rag}~\cite{alvaro2025advanced}.

\section{Conclusion and Future Work}
\label{sec:conclusion}

This paper introduced a~\gls{cca} architecture, achieving real-time adaptability (\textit{RCh1}), enabling context-aware decision-making based on real-time data (\textit{RCh2}), and relaxing pre-defined operational boundaries by dynamically exploring resource capabilities (\textit{RCh3}). The proposed framework enhanced real-time adaptation and resource flexibility using~\gls{llm}. The case study demonstrated that the \gls{cca} architecture effectively mitigates disruptions, improving system resilience and operational flexibility. The case study results showed the architecture enabled the system to reassign tasks to exploration agents, significantly reducing downtime costs in high-mix manufacturing environments.

Future work will focus on enhancing the transparency and interpretability of \gls{llm}-based decisions to improve system validation and trust in industrial settings. Additional studies will involve evaluating the architecture against more varied and complex disruptions in real-world manufacturing scenarios to assess resilience and effectiveness compared to conventional rule-based adaptation methods. To further ensure the practicality of the proposed system, future research will explore the impact of~\gls{llm} response latency in real-world manufacturing environments. The goal is to demonstrate dependable and efficient performance in highly dynamic and practical manufacturing scenarios.



\balance


\end{document}